\documentclass{kapproc} 

\usepackage{procps}

\usepackage[dvips]{graphicx}
\usepackage{wrapfig}

\upperandlowercase

 \let\footnote\savefootnote

\normallatexbib 

\newcommand{\msolar}{\mbox{\,$M_{\odot}$}}

\begin{document}

\articletitle {3D HD and MHD Adaptive Mesh Refinement Simulations
of the Global and Local ISM}

\chaptitlerunninghead{3D HD and MHD AMR Simulations of the ISM} 

\author{ Miguel A. de Avillez\altaffilmark{1} and Dieter Breitschwerdt\altaffilmark{2}}

\affil{\altaffilmark{1}Department of Mathematics, University of \'Evora,
R. Rom\~ao Ramalho 59, 7000 Evora, Portugal, Email:
\texttt{mavillez@galaxy.lca.uevora.pt}\\
\altaffilmark{2}Institut f\"ur Astronomie, Universit\"at Wien,
T\"urkenschanzstr.\ 17, A-1180 Wien, Austria, Email:
\texttt{breitschwerdt@astro.univie.ac.at}}



\begin{abstract}
We have performed high resolution 3D simulations with adaptive
mesh refinement, following the ISM evolution in a star forming
galaxy both on small (<1 pc) and large (>10 kpc) scales, enabling
us to track structures in cooling shock compressed regions as well
as the entire Galactic fountain flow. It is shown in an MHD run
that the latter one is not inhibited by a large scale disk
parallel magnetic field. The fountain plays a vital r\^ole in
limiting the volume filling factor of the hot gas. Contrary to
classical models most of the gas between 100K and 8000 K is found
to be thermally unstable. On scales of superbubbles we find that
the internal temperature structure is rather inhomogeneous for an
old object like our Local Bubble, leading to low O{\sc vi} column
densities, consistent with observations.
\end{abstract}


\section{Introduction}

In their seminal paper of a three-phase model regulated by
supernova explosions in an inhomogeneous medium, \cite{mo1977}
predicted a volume filling factor of the hot intercloud medium
(HIM) of {\bf $f_{v, hot} \simeq 0.7 - 0.8$}.  However,
observations point to a value of $\sim 0.5$ (e.g., \cite{de1992})
or even lower when external galaxies are taken into account (e.g.,
\cite{bb1986}).  A way out has been suggested by \cite{ni1989} by
the so-called chimney model, in which hot gas can escape into the
halo through tunnels excavated by clustered supernova (SN)
explosions. Indeed X-ray observations of several nearby edge-on
galaxies have revealed extended, galaxy-sized halos (e.g.,
\cite{w2001}). The transport of gas into the halo is, however,
still controversial, since break-out may be inhibited by a
large-scale disk parallel magnetic field (see e.g.,
\cite{mss1993}). However, \cite{to1998} has performed 3D MHD
simulations of expanding superbubbles, including radiative
cooling, and finds that bubble confinement only occurs when the
energy injection rate is below a critical value of $\sim 10^{37}
\, {\rm erg} \, {\rm s}^{-1}$ (see also \cite{mm1988}) and/or the
field scale height is infinite, which is unrealistic.

Attempts to determine the occupation fraction of the different
phases, and in particular of the hot gas, by means of modelling
the effects of SNe and SBs in the ISM, have been carried out by
several authors (e.g., \cite{fe1995, fe1998, ko1999}).
However, these models do not include the circulation of gas
between the disk and the \emph{full} halo, thus being unable to
resolve the high-$z$ region; neither do they take into account the
mixing between the different phases. Therefore, an estimate of the
volume filling factors may be misleading.
Using the 3D supernova-driven ISM model of \cite{av2000}
incorporating magnetic fields and the adaptive mesh refinement
technique in HD \cite{av1998} and MHD (using a modified version of
\cite{balsara2001}) algorithms coupled to a 3D parallel
(multi-block structured) scheme, we explored the effects of the
establishment of the disk-halo-disk circulation and its importance
for the evolution of the ISM in disk galaxies both with and
without magnetic fields. In this paper we review some of the
results from these simulations (\S3), and compare in the case of
the Local Bubble derived O{\sc vi} column densities with
observations (\S4), followed by a discussion of the dynamical
picture that emerges from these simulations. Other important
issues like the volume filling factors of the ISM ''phases'', the
dynamics of the galactic fountain, the conditions for dynamical
equilibrium and the importance of convergence of these results
with increasing grid resolution, the variability of the magnetic
field with density, the importance of ram pressure in the ISM, and
the amount of gas in the unstable regimes have been treated
elsewhere \cite{av2000, ab2004a, ab2004b, ab2004c, ab2004d}.

\section{Model and Simulations}

We ran HD and MHD simulations of the ISM on scales of kpc, driven
by SNe at the Galactic rate, on a Cartesian grid of $0\leq
(x,y)\leq 1$ kpc size in the Galactic plane and $-10\leq z \leq
10$ kpc in the halo with a finest adaptive mesh refinement
resolution of 0.625 pc (for the HD run) and 1.25 pc (for the MHD
run) starting from a resolution of 10 pc, using a modified version
of the 3D model of \cite{av2000}, fully tracking the
time-dependent evolution of the large scale Galactic fountain for
a time sufficiently long so that the memory of the initial
conditions is completely lost, and a global dynamical equilibrium
is established.

The model includes the gravitational field provided by
the stellar disk, radiative cooling (assuming an optically thin gas in
collisional ionization equilibrium) with a temperature cut off at 10
K, and uniform heating due to starlight varying with $z$. In the
Galactic plane background heating is chosen to initially balance
radiative cooling at 9000 K. With the inclusion of background heating
the gas at $T<10^{4}$ K becomes thermally bistable. The prime sources
of mass, momentum and energy are supernovae types Ia, Ib+c and II with
scale heights, distribution and rates taken from the literature. OB
associations form in regions with density $n\geq 10$ cm$^{-3}$ and
temperature T$\leq 100$~K.  The number, masses and main sequence life
times ($\tau_{MS}$) of the stars in the association are determined
from the initial mass function (IMF).  At the end of $\tau_{MS}$ the
stars explode. The stars from the OB association that turn into
supernovae in the field have their locations determined kinematically
by attributing to each star a random direction and a velocity at the
time of their formation. The canonical explosion energy is $10^{51}$
erg for all types of SNe. The interstellar gas initially has a density
stratification distribution that includes the cold, cool, warm,
ionized and hot ``phases'' in the Galaxy as described in, e.g.,
\cite{fe1998}. The magnetic field has uniform and random components,
initially given by $B_{u}=(B_{u,0}(n(z)/n_0)^{1/2},0,0)$ and
$B_{r}=0$, respectively, where $B_{u,0}=3~\mu$G is the field strength,
$n(z)$ is the number density of the gas as a function of distance from
the Galactic midplane, and $n_0=1 \, {\rm cm}^{-3}$ is the average
midplane density. This random component of the field is built up
during the first millions of years of evolution as a result of
turbulent motions, mainly induced by SN explosions. The total magnetic field strength at any time $t>0$ is given by $\sqrt{B_{u}^{2}+B_{r}^{2}}> 3$ $\mu$G.

\section{Results}

{\bf Global Evolution.} The initial evolution of the magnetized disk
is similar to that seen in the HD run, that is, the initially
stratified distribution does not hold for long as a result of the lack
of equilibrium between gravity and (thermal, kinetic and turbulent)
pressure during the ``switch-on phase'' of SN activity. As a
consequence the gas in the upper and lower parts of the grid collapses
onto the midplane, leaving low density material in its place. However,
in the MHD run the collapse takes longer due to the opposing magnetic
pressure and tension forces.  As soon as enough supernovae have gone
off in the disk building up the required pressure support, transport
into the halo is not prevented, although the escape of the gas takes a
few tens of Myr to occur. The crucial point is that a huge thermal
overpressure due to combined SN explosions can sweep the magnetic
field into dense filaments and punch holes into the extended warm and
ionized H{\sc i} layers. Once such pressure release valves have been
set up, there is no way from keeping the hot over-pressured plasma to
follow the density gradient into the halo.  As a consequence the
disk-halo-disk duty cycle of the hot gas is fully established, in
which the competition of energy input and losses into the ISM by SNe,
diffuse heating and radiative cooling leads the system to evolve into
a dynamical equilibrium state within a few hundred Myr. This time
scale is considerably longer than that quoted in other papers (e.g.,
\cite{ko1999}, \cite{kb2001}), because in these the galactic fountain
has not been taken into account.

{\bf Summary of Main Results.} The simulations show that:

\quad (i) the highest density gas tends to be confined to shocked
compressed layers that form in regions where several large scale
streams of convergent flow (driven by SNe) occur. The compressed
regions, which have on average lifetimes of 10-15 Myr, are
filamentary in structure, tend to be aligned with the local field
and are associated with the highest field strengths (in the MHD
run), while in the HD runs there is no preferable orientation of
the filaments. The formation time of these high density structures
depends on how much mass is carried by the convergent flows, how
strong the compression and what the rate of cooling of the
regions under pressure are;

\quad (ii) the volume filling factors of the different ISM phases
depend sensitively on the existence of a duty cycle between the disk
and halo working as a pressure release valve for the hot ($T>
10^{5.5}$ K) phase in the disk. The mean occupation fraction of the
hot phase varies from about 15\% for the Galactic SN rate to $\sim
30$\%, for $\sigma/\sigma_{Gal}=4$, and to $52\%$ for
$\sigma/\sigma_{Gal}=16$ (corresponding to a starburst). Overall the
filling factor of the hot gas does not increase with SN rate as much
as may be expected, since due to the evacuation of the hot phase into
the halo through the duty cycle it never exceeds much more than about
half of the disk volume (see also \cite{ab2004a});

\quad (iii) with the magnetic field present and initially
orientated parallel to the disk varying as $\rho^{1/2}$, transport
into the halo is inhibited but not prevented. As a consequence the
hot gas in the disk has a volume filling factor similar to that in
the corresponding HD simulation (i.e., $\leq 20\%$);

\quad (iv) the magnetic field has a high variability and it is
\emph{largely uncorrelated with the density} suggesting that it is driven by inertial motions (which is consistent with the dominance of
the ram pressure - see below), rather than it being the agent
determining the motions. In the latter case the field would not be
strongly distorted, and it would direct the motions predominantly
along the field lines. Therefore the classical scaling law
$B\sim\rho^{1/2}$ according to the Chandrasekhar-Fermi (CF) model
(1953, \cite{cf1953}) does not hold contrary to what has been claimed
by \cite{kb2001}. We suspect that this discrepancy can be explained by
the (200~ pc)$^{3}$ box that these authors have used, centered in the
Galactic midplane with periodic boundary conditions in all the box
faces; thus they have missed completely the disk-halo disk circulation
and did not allow for a global dynamical equilibrium to be established
(see \cite{ab2004d} for a detailed discussion).

\quad (v) Ram pressure controls the flow for $10^{2}<$T$\leq 10^{6}$
K. For T$> 10^{6}$ K thermal pressure dominates, while for T$\leq
10^{2}$ K (thermally stable branch) magnetic pressure takes over. Near
supernovae thermal and ram pressures determine the dynamics of the
flow. The hot gas in contrast is controlled by the thermal pressure,
since magnetic field lines are swept towards the dense compressed
walls. Up to $80\%$ of the mass in the disk is concentrated in the
classical thermally \emph{unstable} regime $10^{2}<$T$\leq 10^{3.9}$ K
with $\sim 60\%$ of the warm neutral medium (WNM) mass enclosed in the
$500\leq $T$\leq 5000$ K gas. This result is in contradiction with the
standard ISM theory \cite{mo1977}, but strongly supported by recent
21cm line observations of the warm neutral medium \cite{heiles03}, who
find a fraction $\geq 48$\% in this unstable temperature range. We
conclude that SN driven shocks and ensuing turbulence are capable of
replenishing gas in the thermally unstable branches by constantly
stirring up the ISM.

\begin{figure}[htbp]
  \centering
Left: mavillez$\_$.fig1a.jpeg \hspace*{1cm}\includegraphics[width=0.5\hsize,angle=0]{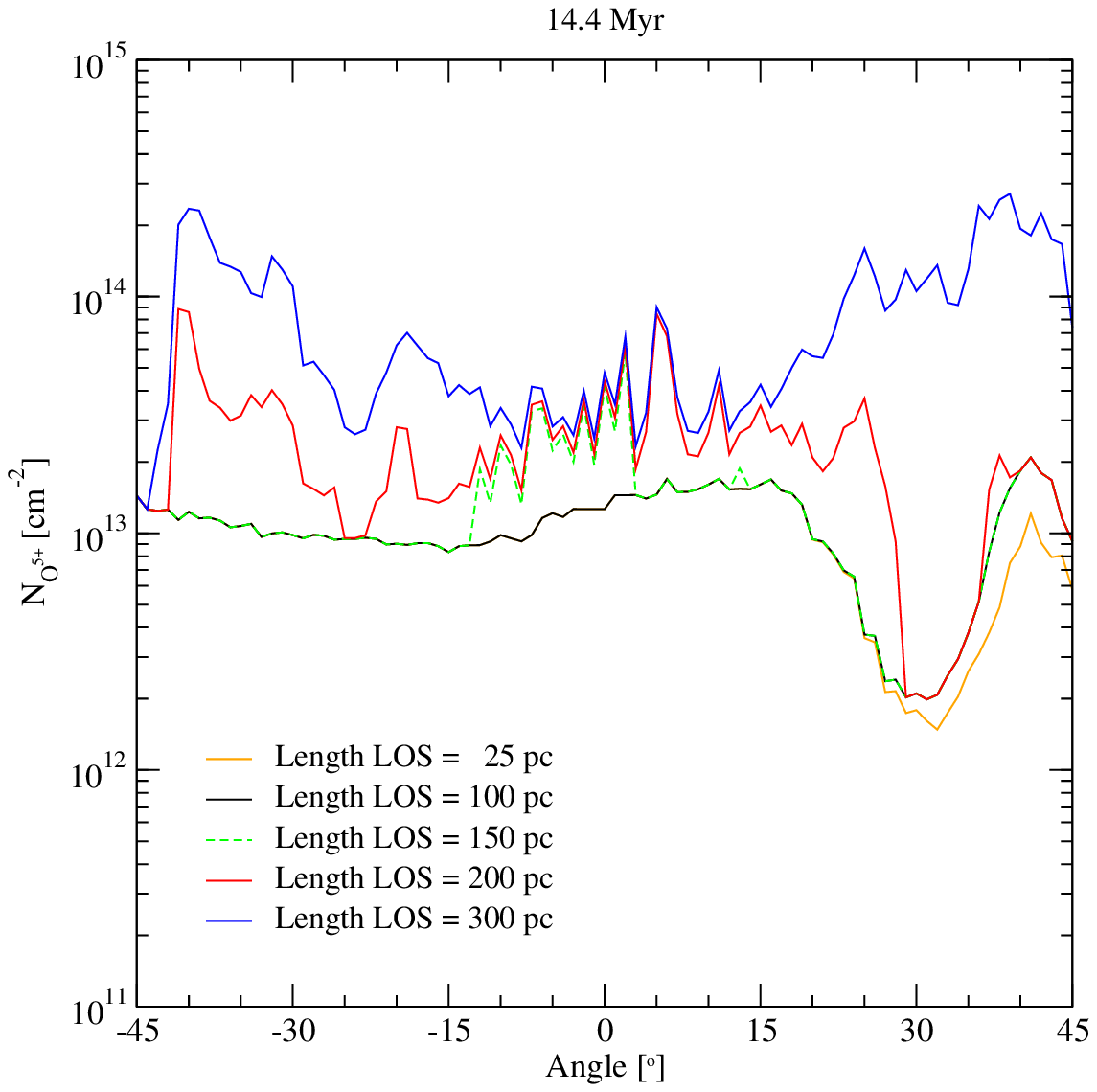}\\
\caption{\emph{Left:} Temperature map (cut through Galactic plane)
of a 3D Local Bubble simulation at 14.4 Myr after
the first explosion showing the LB centered at (175, 400) pc and
Loop I centered at 200 pc to the right along the straight line.
\emph{Right:} O{\sc vi} column density varying over the angles $\pm45^\circ$ for different LOS path lengths. \label{fig1}}
\end{figure}
\section{Does the SN Driven ISM Model hold on smaller scales?}

In the previous sections we have demonstrated how the interstellar
matter cycle works on global scales. How about smaller ones? Let us
look for the remainder of this paper at local superbubbles, i.e. the
Local and Loop I bubbles, and, more specifically, at the O{\sc vi}
column density in absorption, which is a crucial test for modelling
(cf.\ \cite{cox03}). We therefore carried out a run for the Local
Bubble and Loop I following the work by \cite{bb2002}, who suggested
that the LB was created by successive explosions of stars from the
moving subgroup B1 of the Pleiades in the last $15$ Myrs. We used data
cubes of previous runs with a finest AMR resolution of 1.25 pc taken
after the global dynamical equilibrium has been established and
followed the trajectory of the moving group having 20 stars with
masses between 11 and 20 $\msolar$ as predicted by the IMF derived in
\cite{bb2002}.  This run over 20 Myrs with boundary conditions
discussed in \S2, followed the effects of the 20 and 39 SN explosions,
respectively, in the LB and Loop I bubbles. The SNe in the LB went off
along a path crossing the arbitrarily chosen location at $(175,400)$
pc, whereas Loop I is powered by explosions in the Sco Cen
Association, located here at $(375,400)$ pc (left panel of
Fig.~\ref{fig1}). These SNe generated the cavities in which the LB and
Loop I expand. The Galactic supernova rate has been used for the setup
of {\em other SNe} in the remaining disk as a realistic background
medium, in to which the bubbles expand.

\begin{figure}[htbp]
\centering
\includegraphics[width=0.5\hsize,angle=0]{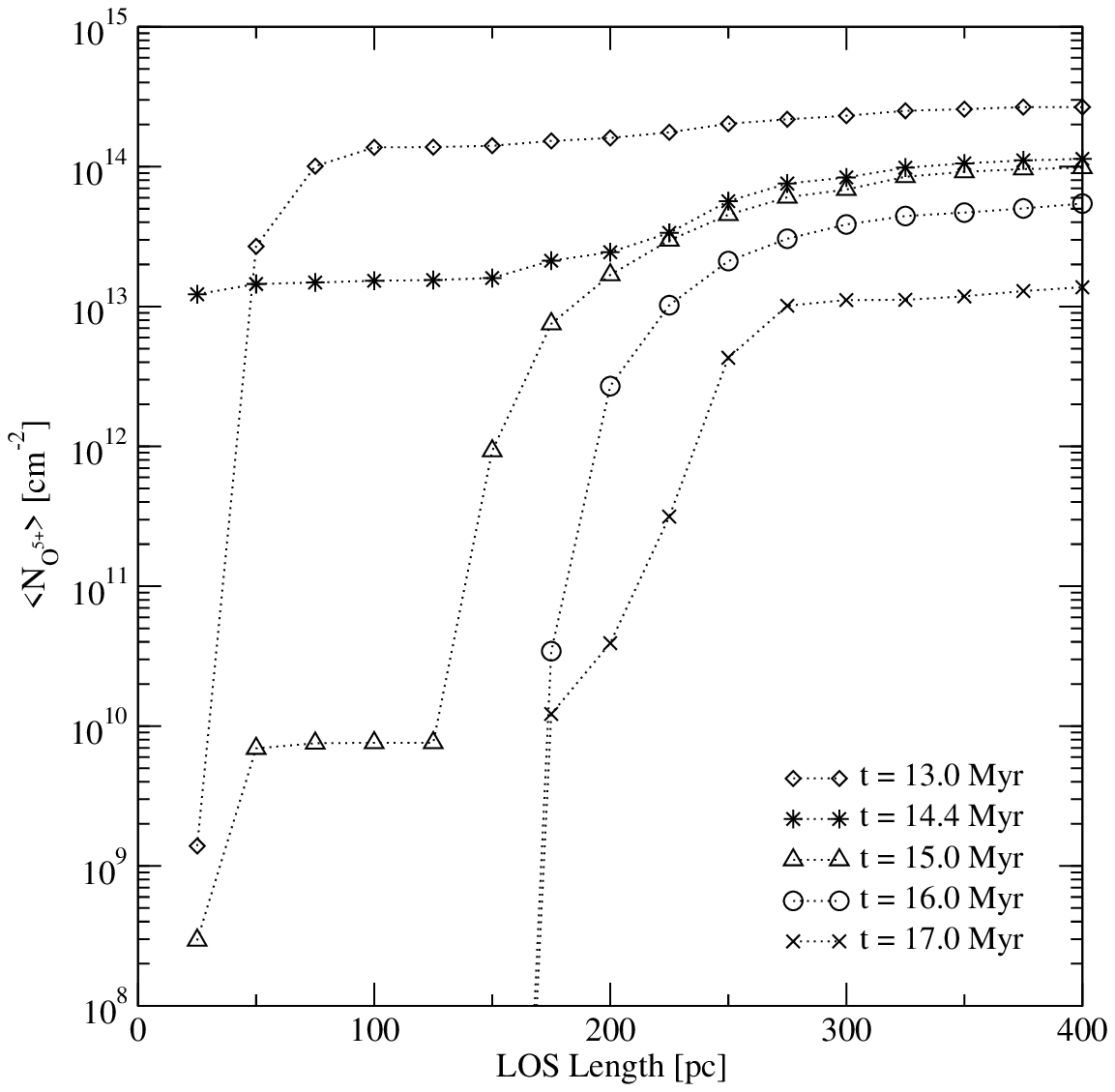}\includegraphics[width=0.5\hsize,angle=0]{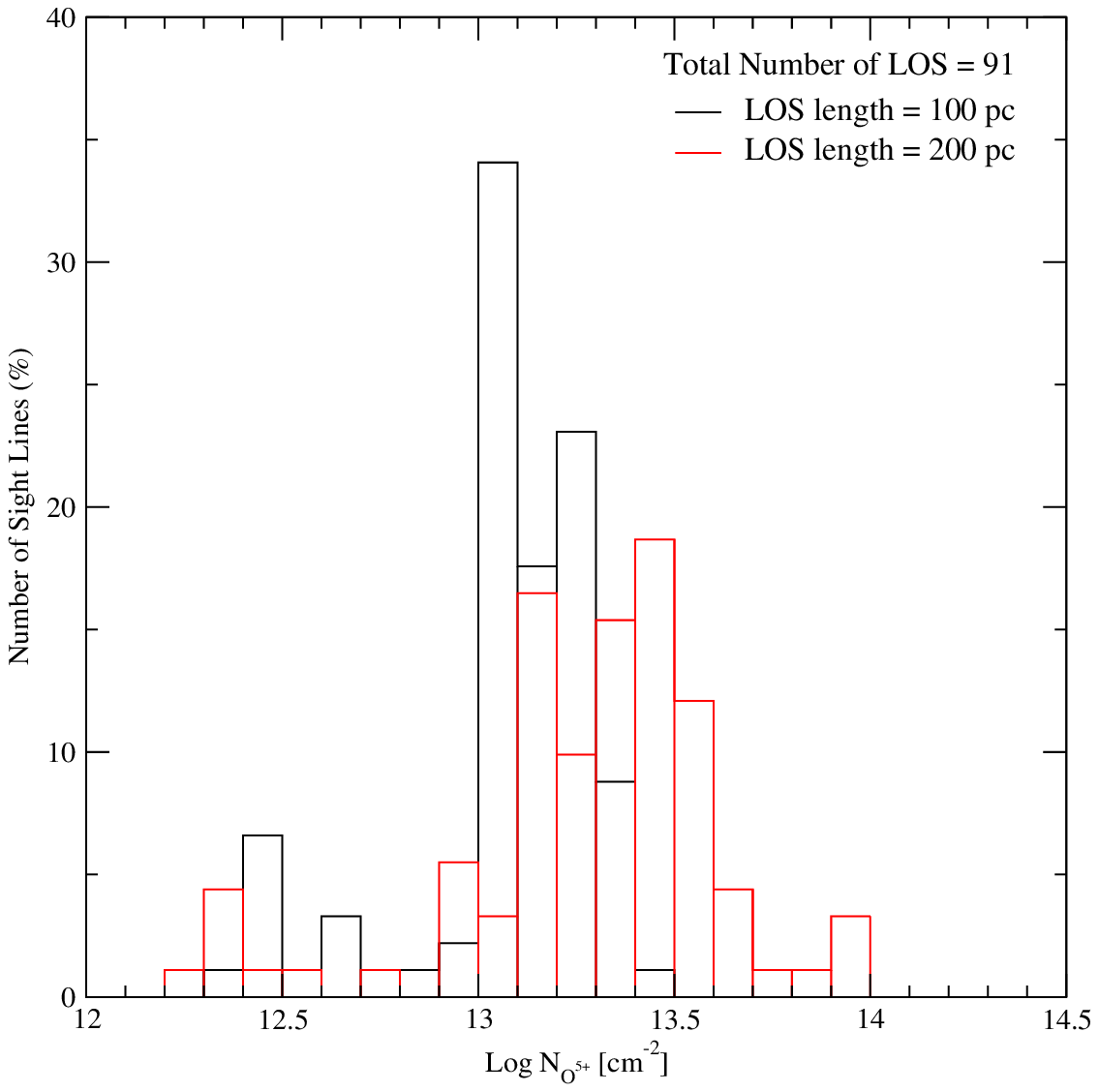}\\
\caption{\emph{Left:} O{\sc vi} column density averaged over
angles indicated in Fig.~\ref{fig1} as a function of LOS path
length at 13, 14.4, 15, 16 and 17 Myr of Local and Loop I bubbles evolution.
\emph{Right:} Histogram showing the percentage of LOS
within various ranges of observed column density for different LOS
path lengths at 14.4 Myr. \label{fig2} }
\end{figure}

The locally enhanced SN rate produces a coherent LB structure within a
highly disturbed background medium (due to ongoing star formation).
Successive explosions heat and pressurize the LB, which at first looks
smooth, but develops internal structure at $t>8$ Myr. After 14 Myr the
20 SNe that occurred inside the LB fill a volume roughly corresponding
to the present day LB (Fig.~\ref{fig1}).  The LB is still bounded by a
shell, which will start to fragment due to Rayleigh-Taylor
instabilities in $\sim 3$ Myr from now. This will lead to mass
transfer of hot gas from Loop I to the Local Bubble, and in $\sim 10$
Myr the bubbles will merge. Clouds and cloudlets of various sizes are
formed when the dense shells of the bubbles collide, as has been
predicted by \cite{bfe2000}. The volume filling factor of the hot ambient gas
in this run is still moderately low ($\sim 18\%$).

The right panel of Fig.~\ref{fig1} shows the O{\sc vi} column density
measured through lines of sight (LOS) with different lengths that run
from $0^\circ$ (along $x$-axis) to $\pm 45^\circ$ (pos.  angles
counter\-clock\-wise) as shown in the left panel of the same figure.
These LOS all cross or point towards Loop I (the hot pressured region
located to the right of the LB).  The column density within the LB
varies between $10^{12}$ and $2.1\times 10^{13}$ cm$^{-2}$, while for
LOS sampling gas from outside the LB (i.e., ahead 100 pc) the column
density is in the range $10^{13}$ cm$^{-2}$ and $3\times 10^{14}$
cm$^{-2}$. The average column density of O{\sc vi}
($\langle$N$_{\mbox{O}^{5+}}$$\rangle$) for 14.4 Myr varies between
$1.5\times10^{13}$ cm$^{-2}$ and $1.4\times 10^{14}$ cm$^{-2}$ for a
LOS length $l_{LOS}$ of 100 and 400 pc, respectively (left panel of
Fig.~\ref{fig2}). Within the LB $\langle$N$_{\mbox{O}^{5+}}$$\rangle$
decreases steeply with time for $t> 14.4$ Myr, because no further SN
explosions occur. The right panel of Fig.~\ref{fig2} shows the
histogram of the percentage of lines of sight within various ranges of
observed column density for different $l_{LOS}$. For $l_{LOS}=100$ pc
there are two strong peaks: one at $1.3\times 10^{13}$ cm$^{-2}$ and
another at $1.6\times 10^{13}$ cm$^{-2}$ from absorbing gas inside the
LB. This second peak is consistent with the fact that the main
contribution for the O{\sc vi} column density comes from the LB as
discussed by \cite{sc94}, who inferred an average value of $1.6\times
10^{13}$ cm$^{-2}$ from analysis of \textsc{Copernicus} absorption
line data.

\section{Conclusions}

The dynamical picture that emerges from these simulations is that
the evolution of the ISM in disk galaxies is intimately related to
the vertical structure of the thick gas disk and to the energy
input per unit area by supernovae. The system evolves towards a
dynamical equilibrium state on the global scale if the boundary
conditions vary only in a secular fashion. Such an equilibrium is
determined by the input of energy into the ISM by SNe, diffuse
heating, the energy lost by radiative and adiabatic cooling and
magnetic compression, and is only possible \emph{after} the full
establishment of the Galactic fountain, which for the Milky Way
takes about 300 Myr (\cite{ab2004a}, see also \cite{kahn81}). It
should be emphasized, since disk and halo are dynamically coupled
not only by the escape of hot gas, but also by the fountain
\emph{return} flow striking the disk, that the disk equilibrium
will also suffer secular variations (see also \cite{rb1995}).

Furthermore, the ISM in the disk is dominated by thermal pressure
gradients mostly in the neighborhood of SNe, which drive motions whose
ram pressures are dominant over the mean thermal pressure (away from
the energy sources) and the magnetic pressure. The magnetic field is
only dynamically important at low temperatures, but can also weaken
gas compression in MHD shocks and hence lower the energy dissipation
rate. The thermal pressure of the freshly shock heated gas exceeds the
magnetic pressure by usually more than an order of magnitude and the
B-field can therefore not prevent the flow from rising perpendicular
to the galactic plane.  Thus, the hot gas is fed into the galactic
fountain at almost a similar rate than without field.

On the scales of superbubbles, it is found that their expansion into a
highly turbulent and inhomogeneous medium leads to considerable
deviations from the classical model by developing internal temperature
and density structure for older bubbles. Thus the O{\sc vi} column
densities we find there are fairly low -- and in agreement with
observations -- while to our knowledge other Local Bubble models so
far have failed this test.

\begin{acknowledgments}
  MAdeA is partly supported by the Portuguese Science Foundation (FCT)
  through the FAAC fund. This work has been partially funded by FCT under the project PESO/P/PRO/40149/2000 to MAdeA and DB.
\end{acknowledgments}

\end{document}